\documentclass[aps,prl,reprint,superscriptaddress,showpacs]{revtex4-1}

\usepackage[]{hyperref} 
\usepackage{amsmath}
\usepackage{amsfonts}
\usepackage{textcomp}
\usepackage[]{graphicx}
\usepackage{epstopdf}
\usepackage{mathrsfs}
\usepackage{amsbsy}
\usepackage{cleveref}

\crefformat{equation}{Eq.~(#2#1#3)}
\crefformat{figure}{Fig.~#2#1#3}

\Crefformat{equation}{Equation~(#2#1#3)}
\Crefformat{figure}{Figure~#2#1#3}

\begin{document}
	
	\title{Nonlinear parity-time-symmetric transition in finite-size optical couplers}
	
	\author{Wiktor Walasik}
	\affiliation{Department of Electrical Engineering, University at Buffalo, The State University of New York, Buffalo, New York 14260, USA}
	\author{Chicheng Ma}
	\affiliation{Department of Electrical Engineering, University at Buffalo, The State University of New York, Buffalo, New York 14260, USA}
	\author{Natalia M. Litchinitser}
	\affiliation{Department of Electrical Engineering, University at Buffalo, The State University of New York, Buffalo, New York 14260, USA}
	\email[]{wiktorwa@buffalo.edu}
	
	\date{\today}

	\begin{abstract}
		Parity-time-symmetric ($\mathcal{PT}$-symmetric) optical waveguide couplers offer a great potential for future applications in integrated optics. Studies of nonlinear $\mathcal{PT}$-symmetric couplers present new possibilities for ultracompact configurable all-optical signal processing. Here, we predict nonlinearly triggered transition from a full to a broken $\mathcal{PT}$-symmetric regime in finite-size systems described by smooth permittivity profiles and, in particular, in a conventional discrete waveguide directional coupler configuration with a rectangular permittivity profile. These results suggest a practical route for experimental realization of such systems. 
	\end{abstract}

\maketitle

One of the fundamental assumptions of quantum mechanics states that operators corresponding to physical observables possess a real spectrum of eigenvalues~\cite{Shankar94}. The most common class of operators having purely real eigenvalues are Hermitian operators. However, in 1998 Bender \textit{et al.}~\cite{Bender98,Bender02} showed that a more general class of non-Hermitian  parity-time-symmetric ($\mathcal{PT}$-symmetric) operators also possesses a real spectrum. 
Consequently, Hamiltonians with complex potentials~$V$ that fulfill the condition $V(\mathbf{r}) = V^*(-\mathbf{r})$ have real eigenvalues and describe physical phenomena.
While the realization of such complex potentials in quantum mechanics is challenging, it is much easier to control them in optics. Indeed, the light propagation can be described by an optical analogue of the Schr\"{o}dinger equation, where the complex dielectric permittivity distribution ${\epsilon}(\textbf{r})$ plays the role of potential and the imaginary part of the permittivity corresponds to gain or loss~\cite{El-Ganainy07,Klaiman08}.

The optical $\mathcal{PT}$ symmetry was experimentally demonstrated in linear optical waveguide couplers~\cite{Ruter10}, which are important components for future fast, ultracompact, and configurable all-optical signal processing.  Optical couplers with gain and loss were also studied in the nonlinear regime~\cite{Chen92}, where unidirectionality~\cite{Ramezani10} and suppression of time reversal~\cite{Sukhorukov10} were shown. 
Phenomena related to the nonlinear $\mathcal{PT}$ symmetry were also studied in periodic systems, including solitons~\cite{Musslimani08,Suchkov11,Abdullaev11,Alexeeva12,Miri12,Wang13}, breathers~\cite{Barashenkov12} and their stability~\cite{Nixon12,Zezyulin12}.

Optical $\mathcal{PT}$-symmetric systems, where ${\epsilon}(\textbf{r}) = {\epsilon}^*(-\textbf{r})$, undergo a transition with the change of the ratio between the imaginary $\epsilon_{\textrm{IM}}$ and the real part $\epsilon_{\textrm{RE}}$ of the permittivity modulation depth ($\epsilon_{\textrm{IM}}/\epsilon_{\textrm{RE}}$). Below a certain threshold value of this ratio, the system is in the full $\mathcal{PT}$-symmetric regime and has purely real eigenvalues~\cite{Guo09,Ruter10}. 
For the $\epsilon_{\textrm{IM}}/\epsilon_{\textrm{RE}}$ ratio above this threshold value, the system is in the broken $\mathcal{PT}$-symmetric regime. In this regime, a pair of modes possesses effective indices that are a pair of complex conjugate numbers, such that one mode experiences gain and the other loss. In linear systems, the transition from the full to the broken $\mathcal{PT}$-symmetric regime (hereafter simply $\mathcal{PT}$~transition) is usually controlled by changing the amount of gain and loss in the system (varying $\epsilon_{\textrm{IM}}$). Nevertheless, the ratio $\epsilon_{\textrm{IM}}/\epsilon_{\textrm{RE}}$ is also influenced by the amplitude of $\epsilon_{\textrm{RE}}$, which in nonlinear systems can be controlled by varying the incident light intensity. Recently, it has been shown that the nonlinearity can trigger a $\mathcal{PT}$ transition in an infinite periodic array of $\mathcal{PT}$-symmetric waveguides described by cosine-like permittivity distribution~\cite{Lumer13}. 

\begin{figure}[!t]
	\includegraphics[width = \columnwidth, clip=true, trim = {0 0 0 0}]{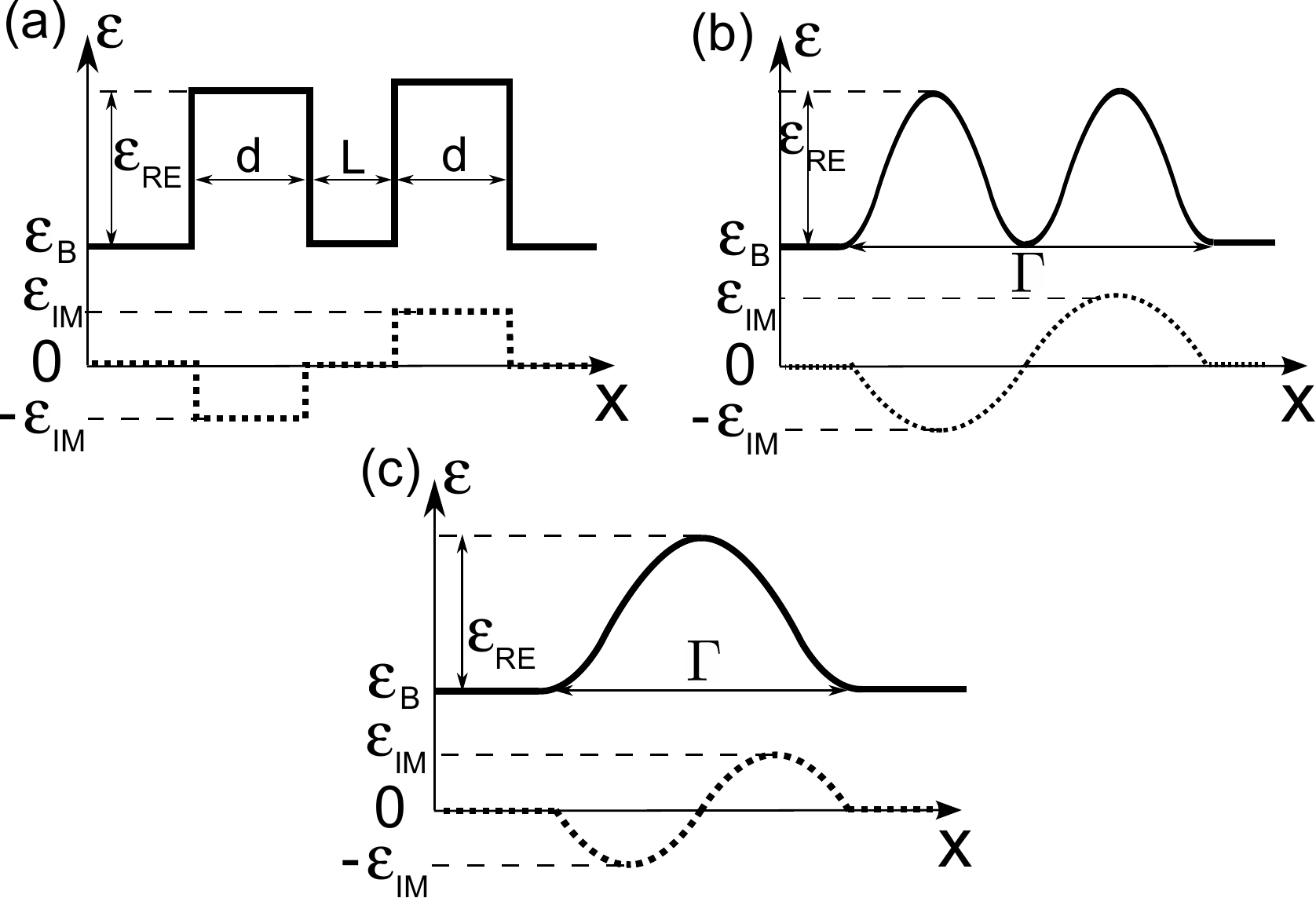}
	\caption{Geometry of the studied structures with their parameters for different types of $\mathcal{PT}$ symmetric couplers. The size of rectangular waveguides is denoted by $d$, and the distance between them is $L$. The size (period) of the cosine-like dimers is denoted by $\Gamma$. $\epsilon_B$ denotes the background relative permittivity; $\epsilon_{\textrm{RE}}$ and $\epsilon_{\textrm{IM}}$ denote the modulation amplitude of the real and the imaginary parts of relative permittivity, respectively.}
	\label{fig:geom}
\end{figure}

In this paper, we show that the nonlinear $\mathcal{PT}$~transition can be observed in ultracompact nonlinear couplers consisting of one or two $\mathcal{PT}$-symmetric waveguides, which can be readily integrated on a chip. 
\Cref{fig:geom} shows the three geometries under investigation: $\mathcal{PT}$-symmetric couplers built of rectangular (a) and sinusoidal (b) waveguides; and a dimer (c), where both gain and loss are located in one waveguiding structure.

To study the nonlinear light propagation in one-dimensional (1D) $\mathcal{PT}$-symmetric structures with Kerr-type nonlinearity, we use the scalar wave equation for the electric field $E(x,z)$:
\begin{equation}
\left[\nabla^2  + k_0^2 {\epsilon}(x) + k_0^2 \alpha(x) |E|^2\right] E = 0,
\label{eqn:wave}
\end{equation}
where the operator $\nabla^2 = \partial^2/\partial x^2 + \partial^2/\partial z^2$ denotes the 2D Laplacian operator, $k_0 = 2\pi/\lambda$ is the free-space wavevector, and $\lambda$ is the free-space wavelength of light. The light propagates along the $z$-direction, and both the structure and the field distributions are assumed to be invariant along the $y$-direction. The linear relative complex permittivity distribution along the $x$-coordinate is described by ${\epsilon}(x) = \epsilon_B + \Delta \epsilon(x)$, where $\epsilon_B$ denotes the background relative permittivity and $\Delta \epsilon(x)$ describes the linear complex permittivity modulation depth. The Kerr nonlinearity strength is quantified by 
$\alpha(x) = \epsilon_0 \Re e\{{\epsilon}(x)\} c n_2(x)$,
where $\epsilon_0$ denotes the vacuum permittivity, $c$ denotes the speed of light, and $n_2$ is nonlinear parameter corresponding to cubic nonlinearity used in the definition of the intensity dependent refractive index $\Delta n_{\textrm{NL}} = n_2 I$~\cite{Boyd07}. The light intensity is related to the electric field in the following way: $I = \epsilon_0 n_B c|E|^2/2$, where $n_B = \sqrt{\epsilon_B}$. The nonlinearity $\alpha(x)$ in the structures studied here is nonzero only inside of the waveguides [i.e., if ${\epsilon}(x) = \epsilon_B $, then $\alpha(x) =  0$].

Using the slowly varying envelope approximation $E(x,z) = \psi(x,z) e^{-i k_0 n_B z}$, \cref{eqn:wave} is transformed into the $(1+1)$D nonlinear Schr\"{o}dinger equation
\begin{equation}
\frac{\partial \psi}{\partial z} = -\frac{i}{2 n_B} \left[ \frac{1}{k_0} \frac{\partial^2}{\partial x^2} + k_0\left( \Delta {\epsilon} + \alpha |\psi|^2\right)\right] \psi.
\label{eqn:NLSE}
\end{equation}
\Cref{eqn:NLSE} is solved using the split-step Fourier method \cite{Feit78,Lax81} in order to analyze the nonlinear dynamics of the light propagation in our $\mathcal{PT}$-symmetric structures.

%
%


While some of the previous theoretical studies focused on the nonlinear light propagation in an infinite periodic cosine-like $\mathcal{PT}$-symmetric waveguide array~\cite{Lumer13}, here we show that the rich variety of nonlinear phenomena can be obtained in simpler finite-size systems of waveguides that are more feasible from the laboratory viewpoint. 
First, we study a nonlinear coupler built of two rectangular waveguides presented in \cref{fig:geom}(a) and look for the nonlinear transition from the full to the broken $\mathcal{PT}$-symmetric regime. The structure with the following parameters is studied: $d = 1$~$\mu$m, $L = 0.1$~$\mu$m, $\epsilon_B = 2$, and $\epsilon_{\textrm{RE}} = 0.02$ at the wavelength $\lambda = 0.63$~$\mu$m.

\begin{figure}[!t]
	\includegraphics[width = 0.49\columnwidth, clip=true, trim = {0 0 0 0}]{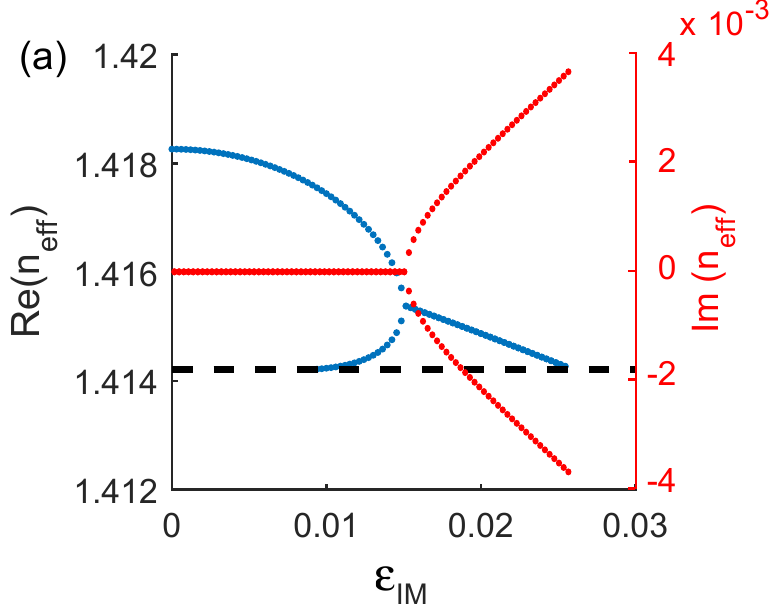}
	\includegraphics[width = 0.49\columnwidth, clip=true, trim = {0 0 0 0}]{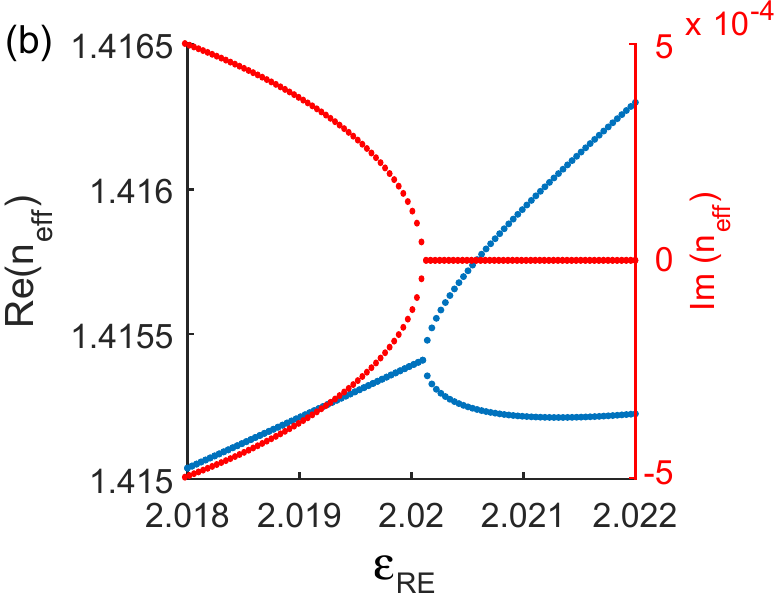}
	\includegraphics[width = 0.49\columnwidth, clip=true, trim = {0 0 20 0}]{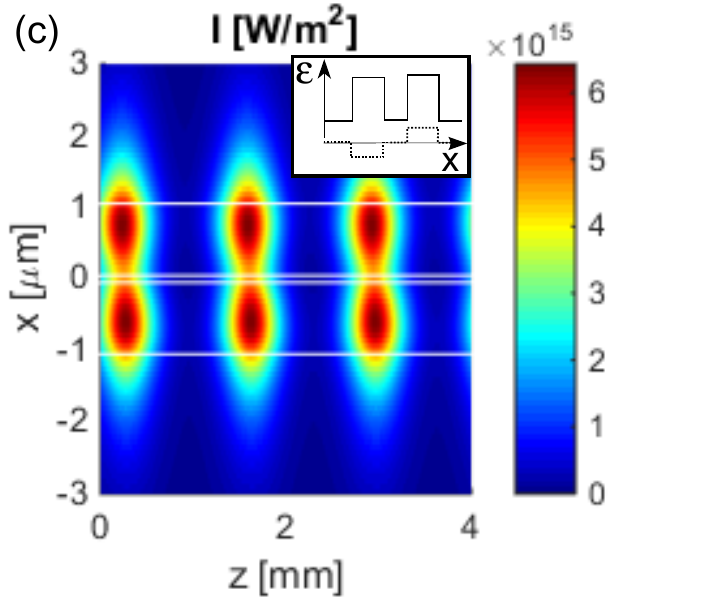}
	\includegraphics[width = 0.49\columnwidth, clip=true, trim = {0 -5 15 5}]{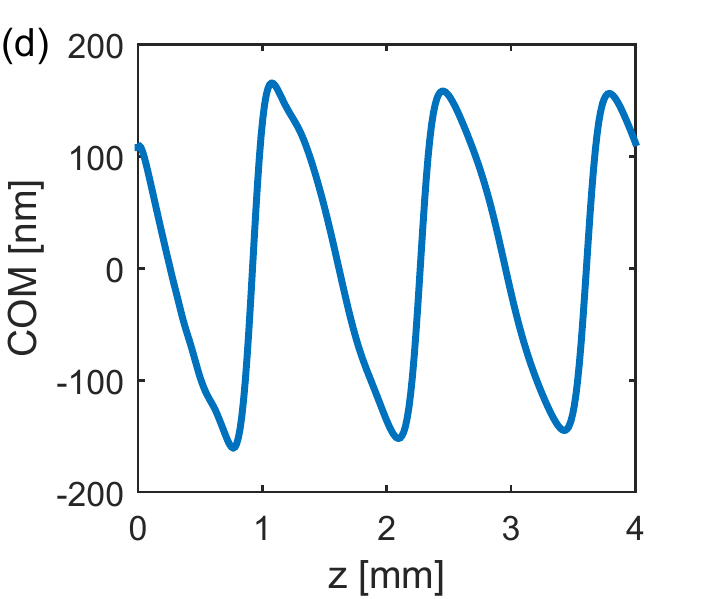}
	\includegraphics[width = 0.49\columnwidth, clip=true, trim = {0 0 20 0}]{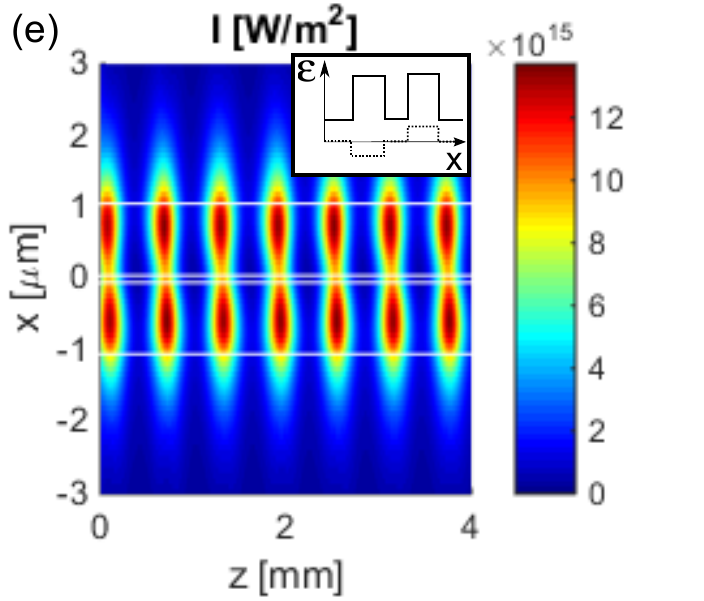}
	\includegraphics[width = 0.49\columnwidth, clip=true, trim = {0 0 20 0}]{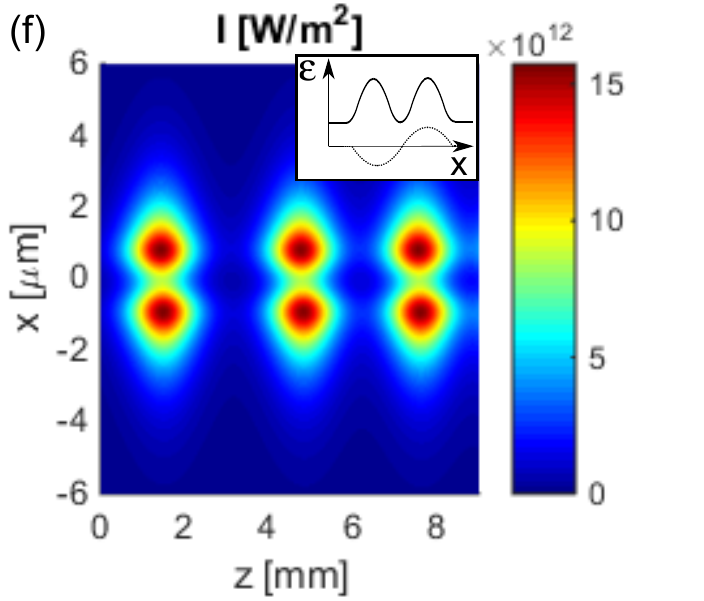}
	\caption{(a),~(b) Linear dispersion relations for a pair of rectangular waveguides shown in \cref{fig:geom}(a). The real (blue curves) and imaginary (red curves) parts of the effective index are shown as a function of the (a) imaginary $\epsilon_{\textrm{IM}}$  and (b) real $\epsilon_{\textrm{RE}}$ part of the permittivity modulation depth. For (a) $\epsilon_{\textrm{RE}}$ is fixed and equal to $0.02$, whereas for (b) $\epsilon_{\textrm{IM}}$ is fixed end equal to $ 0.015$. The black dashed line denotes $n_B$.
		(c)--(f)~Nonlinear dynamics of light propagating in dimers built of (c)--(e) two rectangular waveguides (where  $n_2 = 10^{-19}$~m$^2$/W) and (f) two cosine-like waveguides (where $\alpha = 10^{-19}$~m$^2$/V).
		(c),~(e),~(f) The intensity distribution $I(x,z)$ and (d) the evolution of the $x$-coordinate of the center of mass.
		The geometry of the waveguides is presented in \cref{fig:geom}(a),~(b) and in the figure insets.
		The initial power density $P_0$ is set to (c),~(d) $10^{10}$~W/m, (e) $3 \cdot 10^{10}$~W/m, and (f) $5 \cdot 10^{6}$~W/m. 
		White horizontal lines indicate waveguide boundaries of rectangular waveguides.}
	\label{fig:pair_highP}
\end{figure}

In order to study the properties of light in our $\mathcal{PT}$-symmetric structures, we solve \cref{eqn:wave} using two different methods. First, we study modal properties of the system in the linear regime [$\alpha(x) \equiv 0$]. We look for the field profiles that propagate with an effective index $n_{\textrm{eff}}$ without changing their shape in the form $E(x,z) = \phi(x) e^{-i k_0 n_{\textrm{eff}} z}$. This allows us to transform \cref{eqn:wave} into an eigenvalue problem for the modes of the waveguide described by ${\epsilon}(x)$ and find the corresponding dispersion relations.  We use these dispersion diagrams [shown in Figs.~\ref{fig:pair_highP}(a),~(b)] to locate the symmetry breaking threshold, where two modes with purely real effective indices are transformed into a pair of modes with complex conjugate effective indices.


For the coupler parameters given above, this $\mathcal{PT}$ transition occurs at $\epsilon_{\textrm{IM}} = 0.0149$. Here, we study the system slightly above the phase transition (at $\epsilon_{\textrm{IM}} = 0.015$, which corresponds to the broken $\mathcal{PT}$-symmetric regime for low light intensity). 
The nonlinear dynamics shown in Figs.~\ref{fig:pair_highP}(c),~(d) proves that it is possible to observe the nonlinearly triggered $\mathcal{PT}$ transition in the dimer built of two rectangular waveguides [(shown in~\cref{fig:geom}(a)].
In this and all of the other configurations studied here, we excite the gain mode that initially increases its energy. At the beginning of the propagation, the light intensity is low and the system is located in the broken $\mathcal{PT}$-symmetric regime. The light is mostly located in the gain waveguide, as it can be seen in \cref{fig:pair_highP}(d) showing the $x$-coordinate of the center of mass (COM) calculated as $\int_{-\infty}^{+\infty} x I(x)\mathrm{d}x / \int_{-\infty}^{+\infty} I(x)\mathrm{d}x$. With the increase of the light intensity, the real part of the permittivity of the waveguides also increases. Since the $\mathcal{PT}$~transition depends on the ratio between the real and imaginary parts of permittivities~\cite{Lumer13}, the increase of $\epsilon_{\textrm{RE}}$ drives the system back to the full $\mathcal{PT}$-symmetric regime [the dispersion relation $n_{\textrm{eff}}(\epsilon_{\textrm{RE}})$ presented in Fig.~\ref{fig:pair_highP}(b) shows that with the increase of $\epsilon_{\textrm{RE}}$, the system is transformed from the broken to the full  $\mathcal{PT}$-symmetric regime]. After the $\mathcal{PT}$~transition occurs, the light intensity in the gain waveguide reaches its maximum, and then it starts to diminish. At the same time, energy is coupled to the lossy waveguide in which the maximum intensity is reached slightly later and then also starts to decay. With the decrease of the total intensity in the coupler, accompanied by the decrease of the nonlinear permittivity modulation depth, the system is transformed back to the broken $\mathcal{PT}$-symmetric regime. This cycle repeats itself during the propagation.

\Cref{fig:pair_highP}(e) presents the light evolution in the same system as the one in Figs.~\ref{fig:pair_highP}(a),~(b) but for a higher excitation power density $P_0 = \int_{-\infty}^{+\infty} I(x,z=0) \mathrm{d}x$. We can observe that with the increase of the initial power, the period of oscillations decreases and the maximum light intensity increases. At the low power level, the oscillation period is approximately equal to $1.5$~mm, and the peak power corresponds to the nonlinear permittivity modulation depth of the order of $1.7\cdot10^{-3}$. For high power, the period decreases to $0.6$~mm, and the maximum nonlinear permittivity modulation depth increases to $3.5 \cdot10^{-3}$.

Let us now compare these results for the discrete rectangular waveguide-based coupler with those for the coupler with a smooth cosine-like profile described by 
\begin{equation}
\Delta {\epsilon}(x) =  \epsilon_{\textrm{RE}} \cos^2(2\Omega x) + i \epsilon_{\textrm{IM}} \sin[2\Omega(x - \pi/4)]
\end{equation} and shown in~\cref{fig:geom}(b). Here, $\Omega = \pi/\Gamma$, $\Gamma$ denotes the full width of the dimer, and both $\epsilon_{\textrm{RE}}$ and $\epsilon_{\textrm{IM}}$ are real quantities. While such smooth cosine-like profiles are often used in theoretical studies of periodic structures and waveguides, it maybe challenging to realize such profiles in laboratory experiments. The cosine-like coupler parameters are: $\epsilon_B = 2$, $\epsilon_{\textrm{RE}} = 0.03$, and $\Gamma = 3$~$\mu$m. The imaginary part of the permittivity is chosen to be $\epsilon_{\textrm{IM}} =  0.0162$, which is just above the $\mathcal{PT}$~transition point located at $\epsilon_{\textrm{IM}} =  0.016$. It is noteworthy, that the dynamics of nonlinear wave propagation in such a structure, shown in \cref{fig:pair_highP}(f), resembles that of the discrete waveguides based coupler [see \cref{fig:pair_highP}(e)], suggesting that the nonlinearity-induced $\mathcal{PT}$~transition is not sensitive to the exact refractive index profile of the structure. These results are likely to be important for the experimental observation of the predicted phenomena.



Now, having established the similarity between light propagation in $\mathcal{PT}$-symmetric structures defined by rectangular and smooth permittivity profiles, we consider the effect of different distributions of both the real and imaginary parts of the dielectric permittivity on the $\mathcal{PT}$~transition dynamics. In particular, we investigate a finite-size $\mathcal{PT}$-symmetric dimer that can be described by 
\begin{equation}
\Delta {\epsilon}(x) =  \epsilon_{\textrm{RE}} \cos^2(\Omega x) + i \epsilon_{\textrm{IM}} \sin(2 \Omega x),
\label{eqn:cos_ind}
\end{equation}
with Kerr-type nonlinearity. The geometry of a single dimer described by \cref{eqn:cos_ind} is shown in \cref{fig:geom}(c), and the parameters are $\epsilon_B = 2$, $\epsilon_{\textrm{RE}} = 0.05$, and $\Gamma = 3$~$\mu$m.  Here, the transition from the full to the broken $\mathcal{PT}$-symmetric regime is also observed. The threshold in the linear case is located at $\epsilon_{\textrm{IM}} = 0.0305$. 

\begin{figure}[!t]
	\includegraphics[width = 0.49\columnwidth, clip=true, trim = {0 0 20 0}]{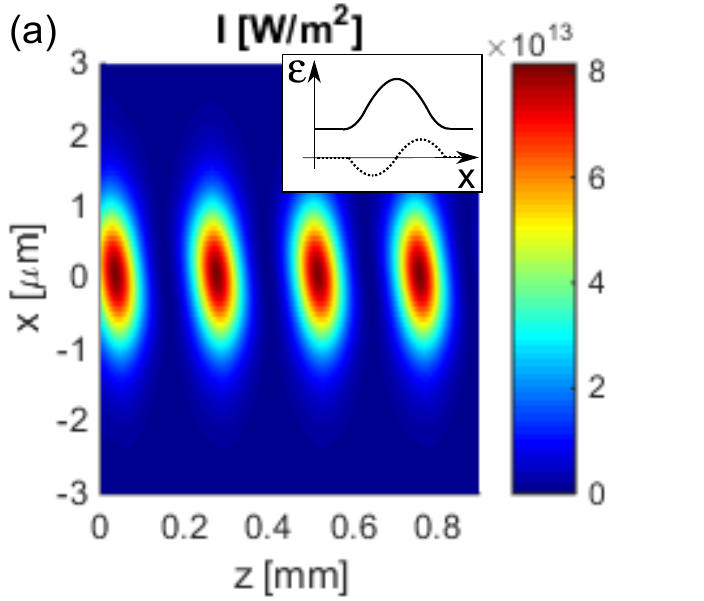}
	\includegraphics[width = 0.49\columnwidth, clip=true, trim = {0 -5 15 5}]{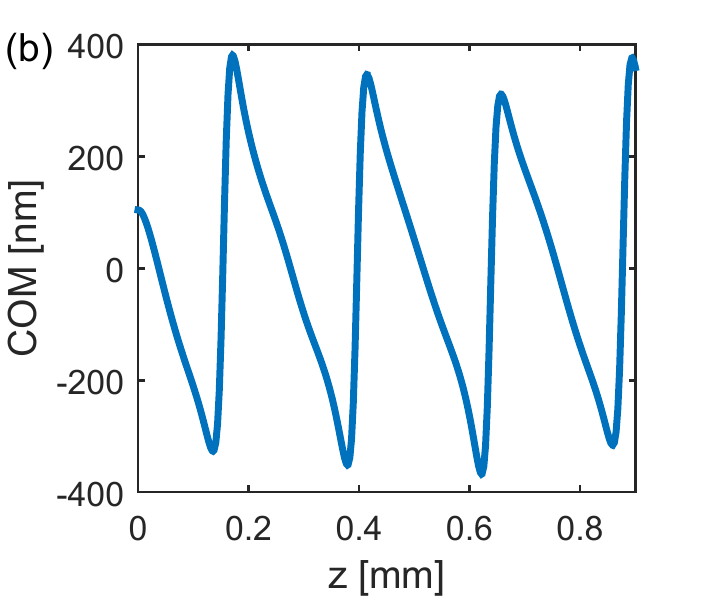}
	\includegraphics[width = \columnwidth, clip=true, trim = {15 0 45 0}]{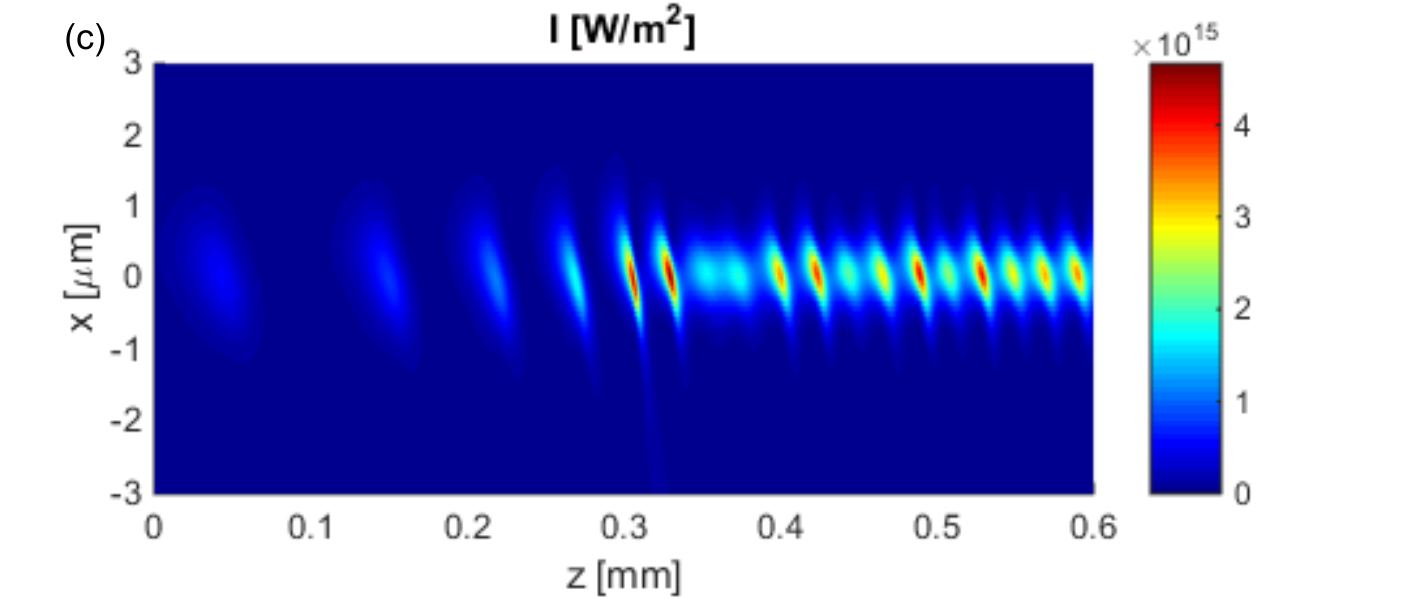}
	\caption{Nonlinear dynamics of light propagating in a single dimer described by \cref{eqn:cos_ind}.
		(a),~(c) The intensity distribution $I(x,z)$, (b) the evolution of the $x$-coordinate of the center of mass corresponding to subplot (a). The geometry of the structure is shown in \cref{fig:geom}(c) and in the inset of supblot (a). The parameters are: $\epsilon_{\textrm{RE}} = 0.05$, $\alpha = 10^{-19}$~m$^2$/V$^2$, initial power density $P_0 = 10^8$~W/m, and the imaginary part of permittivity $\epsilon_{\textrm{IM}}$ equals (a),~(b) $0.031$ and (c) $0.035$.}
	\label{fig:single_per_evo}
\end{figure}

The nonlinear propagation of light in a single dimer is presented in Figs.~\ref{fig:single_per_evo}(a),~(b). It resembles the propagation of light in the unit cell of a periodic array of dimers presented in Figs.~3(d),~(f) in Ref.~\cite{Lumer13}. However, in Ref.~\cite{Lumer13} the parameters of the array were chosen in such a way that the energy transfer from the gain region to the lossy one was visible only from the plot showing the center of mass and not from the color map presenting the intensity evolution during the propagation. On the contrary, for the parameters chosen here, this energy transfer, illustrated as the displacement of the center of mass in \cref{fig:single_per_evo}(b), is also clearly visible from the intensity distribution presented in \cref{fig:single_per_evo}(a). Indeed, the intensity distribution pattern that changes periodically along the $z$-coordinate is visibly tilted away from the vertical axis, indicating the energy transfer from the top part of the dimer (gain region) to the bottom part (loss region).

Figures~\ref{fig:single_per_evo}(a),~(b) show that the system is initially (for low light intensity) in the broken $\mathcal{PT}$-symmetric regime. The gain mode of the system is excited, and the energy of this mode increases. Similar to the case shown in \cref{fig:pair_highP}, this increase is accompanied by the increase of the real part of the permittivity due to the Kerr effect. When the permittivity modulation depth is sufficiently high, it brings the system back to the full $\mathcal{PT}$-symmetric regime. Initially, the energy is stored in the gain mode of the broken $\mathcal{PT}$-symmetric system. In the full $\mathcal{PT}$-symmetric regime, this mode is no longer an eigenmode of the system, and the distribution of light (still experiencing gain as it is localized mostly in the gain region) is attracted to the center of the dimer, where the permittivity is the highest~\cite{Lumer13}. Therefore, the beam possesses a nonzero transverse momentum directed towards negative $x$-direction. Due to this momentum, the beam crosses the center of the dimer and as a result is mostly localized in the region with loss. Consequently, the total energy rapidly decreases and the system is transformed back to the broken $\mathcal{PT}$-symmetric regime. In this regime, the part of the energy stored in the loss mode vanishes quickly and the part stored in the gain mode starts to grow again. This cycle repeats, as it is seen in Figs.~\ref{fig:single_per_evo}(a),~(b). 

It is noteworthy that in the case of a single dimer, for which the parameters are chosen in such a way, that the system is higher above the $\mathcal{PT}$~transition ($\epsilon_{\textrm{IM}} = 0.035$ while it was $0.031$ in the example discussed above), the nonlinear transition also occurs, but it has a different character. The propagation of light in such a system is presented in \cref{fig:single_per_evo}(c). We see that in the initial phase of the propagation (up to 0.3~mm), the oscillations are aperiodic, and the period of oscillations decreases. With the decrease of the oscillation period, the energy is transferred from the gain part of the dimer to the lossy part within a shorter distance. At the beginning of the propagation, the transverse component (toward negative $x$ values) of the beam momentum is low enough so that the beam stays confined in the dimer. During the propagation, this momentum increases and at the propagation distance $x \approx 0.3$~mm, it causes a significant quantity of energy to be outcoupled from the dimer and to propagate toward negative values of $x$. After this point, the propagation pattern becomes more regular: the period of oscillations is kept constant, and only the peak intensity reached during the oscillations changes (the variations of the peak intensity are at the level of 50\%). The irregularity of the propagation in the system located higher above the $\mathcal{PT}$~transition can be explained by the fact that the light intensity here is 20 times larger than that for the system closer to the $\mathcal{PT}$~transition. At the same time, the period of the oscillations is about 4 times smaller, and consequently, the changes in the light intensity are much more abrupt than in the system closer to the $\mathcal{PT}$~transition. These rapid changes cause more irregular behavior of the system high above the $\mathcal{PT}$~transition.

Here, we briefly compare the field distribution in a pair of waveguides shown in \cref{fig:pair_highP} and the single cosine-like dimer shown in \cref{fig:single_per_evo}. The main difference lies in the fact that in the case of the single cosine-like dimer, the field possesses a single intensity maximum whose location is described by the center of mass that shifts from the gain to the loss part of the waveguide, whereas in the pair of waveguides the field distribution has two maxima---one in each waveguide. In the system of two waveguides, it is the ratio between these maximum intensities (or more precisely the integrated power in each of the waveguides) that determines whether gain or loss dominates at a given propagation distance.


In conclusions, we have studied nonlinear dynamics of light propagating in finite-size parity-time-symmetric optical couplers. We have shown that the nonlinearly triggered transition from the full to the broken $\mathcal{PT}$-symmetric regime can be observed for various finite-size couplers built of two rectangular and cosine-like waveguides. We have shown that the nonlinearly triggered transition between the full and the broken $\mathcal{PT}$-symmetric regime also occurs in a single $\mathcal{PT}$-symmetric cosine-like dimer. Moreover, we have found that the light propagation in smooth and discrete waveguide couplers is qualitatively similar. These results suggest a practical and simple route to experimental verification and applications of this kind of nonlinear $\mathcal{PT}$~transition. 

\section*{Acknowledgements}

We thank Dr.~Liang Feng from University at Buffalo, The State University of New York for helpful discussions. This work was supported by US ARO Award \#W911NF-15-1-0152.





%

\end{document}